\documentclass[aps,twocolumn,showpacs,showkeys,preprintnumbers,amsmath,amssymb,floatfix,superscriptaddress]{revtex4}
\usepackage{graphicx}
\usepackage{bm}
\usepackage{float}

\begin{document}


\title{Cooling and trapping of atoms and molecules by the counter-propagating pulses trains}

\author{V. I. Romanenko}
\email{vr@iop.kiev.ua}
\author{Ye. G.Udovitskaya}
\affiliation{Institute of Physics, Nat. Acad. Sci. of Ukraine, 46, Nauky Ave., Kyiv 03650, Ukraine}
\author{A. V. Romanenko}
\affiliation{ Kyiv National Taras Shevchenko University 2, Academician Glushkov Ave, Kyiv 03022, Ukraine}
\author{L. P. Yatsenko}
\affiliation{Institute of Physics, Nat. Acad. Sci. of Ukraine, 46, Nauky Ave., Kyiv 03650, Ukraine}

\date{\today}

\begin{abstract} 
We discuss a possible one-dimensional trapping and cooling of atoms and molecules due to 
their non-resonant interaction with the counter-propagating light pulses trains. 
The counter-propagating pulses form a one-dimensional trap for atoms and molecules,
and properly chosen the carrier frequency detuning from the transition frequency 
of the atoms or molecules keeps the ``temperature'' of the atomic or molecular ensemble
close to the Doppler cooling limit. The calculation by the Monte-Carlo wave function method is carried out for the two-level and three-level schemes of the atom's and the molecule's interaction 
with the field, correspondingly. The discussed models are applicable  to atoms and molecules with almost diagonal Frank-Condon factor arrays. 
Illustrative calculations where carried out for 
ensemble averaged characteristics   
for sodium atoms and SrF molecules in the trap.  Perspective for the nanoparticle light pulses's trap formed
by counter-propagating light pulses trains is also discussed. 
\end{abstract}

\maketitle

\section{Introduction}

Optical cooling and trapping~\cite{Chu98,Coh98,Phi98} is the key stage of the experiments with cold atoms.
Initially continuous laser radiation is used for this purpose, but pulsed laser applications for 
cooling~\cite{Str89,Mol91,Wat96,Ili11,Ili12} and trapping~\cite{Fre95,Goe97,Bal05,Rom11,Yan13} of atoms and molecules are also discussed now.
Laser cooling of atoms by counter-propagated weak laser pulses was investigated in~\cite{Mol91}, 
but possible trapping was not recognized.
The authors of~\cite{Rom13} analyzed the light pulses interaction with atoms for different detunings and found 
that simultaneous cooling and trapping of atoms are possible, provided that the carrier frequency detuning 
from the resonant atom-light interaction is properly chosen. More deep investigation of the cooling trap,
based on the interaction of atoms with counter-propagating laser pulses, is described in~\cite{Rom14},
where the numerical calculations for examples of the time evolution of a sodium atom in the trap where demonstrated.

\begin{figure}[h]
\begin{center}
\includegraphics{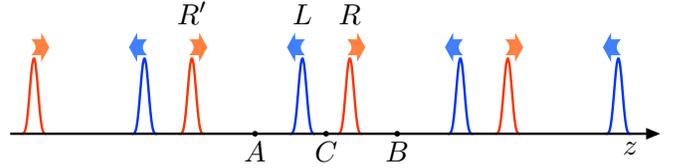}
\caption{Counter-propagating laser pulses form a trap for atoms near point $C$, where the pulses ``collide''}%
\label{figure:1}
\end{center}
\end{figure}

The idea of the trap based on the atom's interaction with the counter-propagating light pulses trains can be most easy explained
for the case of two-level atoms in the field of $\pi$-pulses. Let light pulses propagate along the
$z$-axis (see Fig.~\ref{figure:1}) and point $C$ is the point where the counter-propagating pulses ``collide''.
We assume that an atom at point $A$ was in the ground state before the recent interaction 
with pulse~$R$ (this is true in most cases because
of small time between the interaction of the atom with $R$ and $L$ pulses in comparison with the 
time between the interaction with $R$ and $R'$ pulses~\cite{Voi91}). 
As a result of the interaction with this pulse, the atom absorbed 
a photon and became excited. 
Its momentum was changed by the photon momentum $\hbar{}k$ in the positive $ z $-axis direction.
After being subjected to the action of pulse~$L$, the atom emits a photon, becomes unexcited, 
and its momentum changes by another $\hbar k$ in the same direction.
The interaction of the atom with the laser pulses repeats with period $T$, so the atom is subjected to the action of
the average force $2\hbar k/T$ directed towards point~$C$. A similar reasoning for an atom at  point~$B$ 
allow us to find that the atomic momentum changes by $-2\hbar k$, so that the average force acting on it 
is $-2\hbar k/T$, i.e. directed towards point~$C$. 
From symmetry considerations we readily conclude that the light pressure force on the atom
at point~$C$ equals zero. 
Hence, counter-propagating light $\pi$-pulses can form a
trap for an atom with the center at point $C$, where the counter-propagating pulses ``collide''.  
As was pointed out in \cite{Fre95,Goe97}, pulses with areas different from $\pi$ can also form a trap.

Recently a great progress in the manipulation of molecules by laser radiation was reported~\cite{Bar12}.
The authors of~\cite{Bar12} demonstrated deceleration of a beam of neutral 
strontium monofluoride molecules using radiative
forces. The spectroscopic constants of this molecule satisfies the main conditions, 
which are required for the successful laser cooling.   
They are~\cite{Ros04}: (1) a band system with strong 
one-photon transitions (i.e. large oscillator strength) to ensure the high photon-scattering 
rates needed for rapid laser cooling, (2) a highly-diagonal Franck-Condon array for the band system, and 
(3) no intervening electronic states to which the upper state could radiate and terminate the cycling transition.
We note that it is the violation of the second criterion led to very high (about 97\%) losses 
of the ground working state of 
Na$_{2}$ in the first observation of the light pressure force on molecules~\cite{Voi94}.

In this paper we calculate the characteristics of atomic and molecular ensembles 
in the trap formed by counter-propagating
light pulses using Monte-Carlo method. We apply this method for different purpose: 
(1) simulation of an atom or a molecule states
evolution by the Monte-Carlo wave function (MCWF) method~\cite{Mol93}, and 
(2) calculation of ensemble averages of the coordinate, velocity and the second momenta of their values. 

We illustrate the phenomenon of simultaneous cooling and trapping of atoms and molecules by counter-propagating  
light pulses trains using examples of sodium atoms 
and strontium monofluoride molecules, which have the level structure, suitable 
for light pressure force experiments~\cite{Met99,Bar12}.
We use the two-level model for an atom, as far as it adequately describes 
the cycling cooling transition~\cite{Met99}, and the
three-level $\Lambda$-model for a molecule, as far as 0.9996 of the excited 
molecules radiatively decay to the two 
lower levels~\cite{Bar12}.
The atomic motion is described in the framework of classical mechanics, 
that corresponds to the narrow atomic wave packet in comparison with the wavelength.
A perspective of trapping of nanoparticles is briefly discussed in the final part of the article.

This paper is organized as follows. In Sec.~\ref{sec-mod} we present the the models for atoms 
and molecules used in the paper. 
Section~\ref{sec-pulses} describes the trains of the counter-propagating pulses which acts 
on the atoms and the molecules.
Solving of Schr\"odinger equation by the Monte-Carlo wave function method is described 
in Sec.~\ref{sec-MCWF}, closely following~\cite{Mol93}.
Section~\ref{sec-mech} contains the calculation of light pressure force and equations for mechanical motion of atoms and molecules.
In Sec.~\ref{sec-num} we describe the numerical calculation routine used in the investigation.
Results and discussion are presented in Sec.~\ref{sec-Res}. Short conclusions are formulated in Sec.~\ref{sec-conc}.

\section{Models for atoms and molecules}
\label{sec-mod}

We use the two-level model for description of the atom-field interaction. 
The transitions in atoms, which ensure the cycling interaction
with the field within the two-level system, are listed, for example, in~\cite{Met99}. 
\begin{figure}[ht]
\begin{center}
\includegraphics{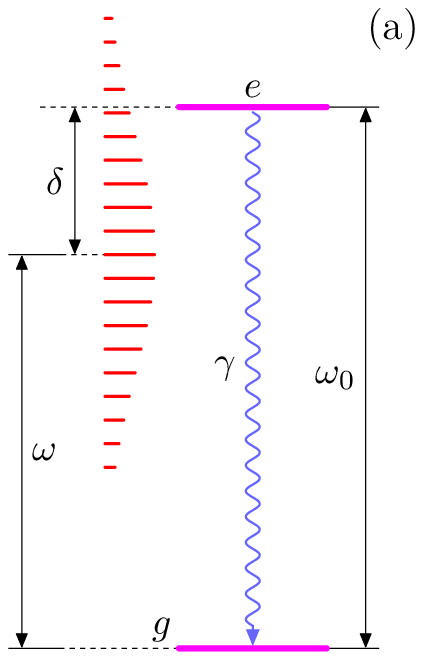}
\bigskip

\includegraphics{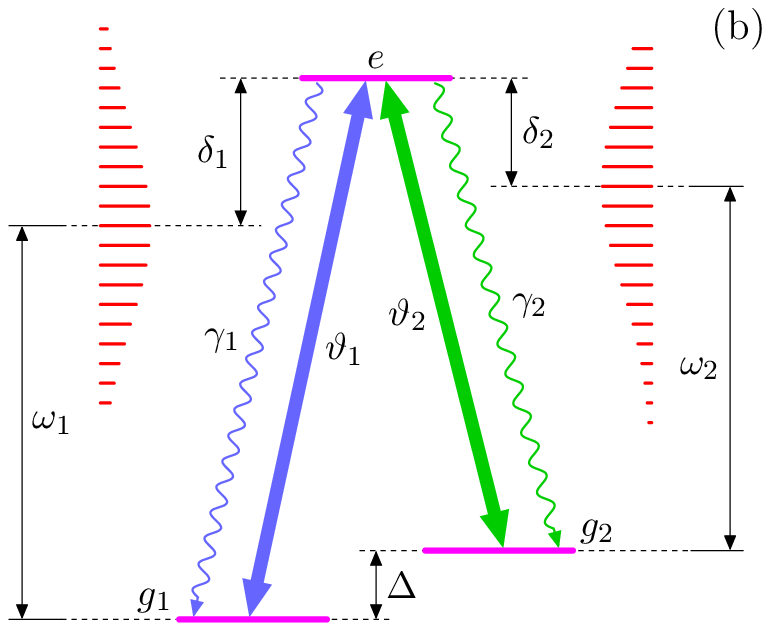}

\caption{Two-level scheme of the atom-field interaction (a) and three-level scheme 
of the molecule interaction with the field of laser radiation (b)}%
\label{figure:2}
\end{center}
\end{figure}
We denote the ground state with $g$ and the
excited state with $e$ [see Fig.~\ref{figure:2}(a)]. The detuning of the carrier frequency $\omega$ from the transition frequency $\omega_{0}$ is 
$\delta=\omega_{0}-\omega$, and the rate of the atom's spontaneous emission from the excited state is $\gamma$.

We describe the molecule's  interaction with the field  by the three-level $\Lambda$-model, 
as depicted in Fig.~\ref{figure:2}(b).
This model  is composed of the excited state $e$ and the ground states $g_{1}$, $g_{2}$, separated by $\hbar\Delta$.
The transition frequencies between the excited and each of the ground states are $\varpi_{1}$,  
$\varpi_{2}$, correspondingly. 
These frequencies differs for SrF molecule, which interaction with the laser pulses is simulated 
in this paper, by $\Delta/2\pi=14.9$~THz~\cite{Bar12}.
As far as $2\pi/\Delta$  is very small in comparison with the pulse duration $\tau$, we need 
two pairs of pulse trains, one of which is the counter-propagating 
pulses close to the resonance with $e\leftrightarrow{}g_{1}$ transition, and the other is the counter-propagating 
pulses close to the resonance with $e\leftrightarrow{}g_{2}$ transition.  The carrier frequency 
of these pulses $\omega_{1}$, $\omega_{2}$  are detuned 
from the resonances by $\delta_{1}=\varpi_{1}-\omega_{1}$ and $\delta_{2}=\varpi_{2}-\omega_{2}$, 
correspondingly. 
We also introduce the spontaneous decay rates $\gamma_{1}$
and $\gamma_{2}$ from the upper state to the two lower states, which form the total decay 
rate $\gamma=\gamma_{1}+\gamma_{2}$. 

\section{Laser pulses}
\label{sec-pulses}

We suppose that the pairs of pulses traveling in the same direction 
(and resonant to the different transitions) coincides in time.
The spectrum of the laser field is a frequency 
comb with the difference between 
the teeth $2\pi/T$, where $T$ is the repetition period of the laser pulses.

The electric field of the trains of the counter-propagating pulses can be written as 
\begin{eqnarray}  
\bm{\mathrm{E}}(t)&=&E_{1}\bm{\mathrm{e}}_{1}\sum\limits_{m}^{}\cos[\omega_{1}{}t-k_{1}z+\varphi_{11,m}]
f(\eta_{1,m})\nonumber\\
&&+E_{1}\bm{\mathrm{e}}_{1}\sum\limits_{m}^{}\cos[\omega{}_{1}t+k_{1}z+\varphi_{12,m}]f(\eta_{2,m})\nonumber\\
&&+E_{2}\bm{\mathrm{e}}_{2}\sum\limits_{m}^{}\cos[\omega_{2}{}t-k_{2}z+\varphi_{21,m}]f(\eta_{1,m})\nonumber\\
&&+E_{2}\bm{\mathrm{e}}_{2}\sum\limits_{m}^{}\cos[\omega_{2}{}t+k_{2}z+\varphi_{22,m}]f(\eta_{2,m}).
\label{eq:E}
\end{eqnarray}
Here $k_{1}=\omega_{1}/c$, $k_{2}=\omega_{2}/c$ are the values of wave vectors for carrier frequencies 
$\omega_{1}$ and $\omega_{2}$,  
$\bm{\mathrm{e}}_{1}$ and $\bm{\mathrm{e}}_{2}$ are polarization vectors,  $\varphi_{11,m}$, $\varphi_{12,m}$ and 
$\varphi_{21,m}$, $\varphi_{22,m}$ are the phases of the counter-propagating $m$-pulses for $t=0$ and $z=0$. 
Function $f(\eta)$ with maximum value $f(0)=1$ describe the shape of the pulse's envelope,
\begin{eqnarray}
\eta_{1,m}&=&\frac{1}{\tau}\left(t-mT-\frac{z}{c}\right),\label{eq:etai}\\
\eta_{2,m}&=&\frac{1}{\tau}\left(t-mT+\frac{z}{c}\right),
\label{eq:etaii}
\end{eqnarray}
where $z$ is the atom's or molecule's coordinate, $\tau$ is the pulse duration. The beginning of the coordinate axis 
and the order of the pulses numbering are chosen 
so that the counter-propagating  pulses number $m$ meet each other at 
time instants $t=mT$ in point $z=0$, where $m$ is an arbitrary integer.                 

The pulse areas are defined by the integrals
\begin{equation}
\vartheta_{j}=\Omega_{j}\int_{-\infty}^{\infty}f(t/\tau)\,dt,\quad{}j=1,2,
\label{eq:theta}
\end{equation}
where the Rabi frequencies are
\begin{equation}
\Omega_{j}=-\bm{\mathrm{d}}_{g_{j}e}\bm{\mathrm{e}}_{j}E_{j}/\hbar,\quad{}j=1,2.
\label{eq:Rabi}
\end{equation}
The matrix elements $\bm{\mathrm{d}}_{g_{j}e}=\langle{g_{j}}|\bm{\mathrm{d}}|e\rangle$ of 
the dipole moments are assumed
to be the real-valued quantities without loss of generality~\cite{Sho90}. 

The case of the two-level model [Fig.~\ref{figure:2}(a)] is described by the equation of 
this subsection with  $\gamma_{1}=\gamma$,
$\gamma_{2}=0$, $\vartheta_{1}=\vartheta$, $\vartheta_{2}=0$, $\omega_{1}=\omega$, 
$\delta_{1}=\delta$, $\varpi_{1}=\omega_{0}$.

Usually  the Gaussian-like pulses are used in simulations of the atom-field interactions~\cite{Ber98,Vit01}. 
These pulses are artificially cut off beyond certain limits in numerical calculation.
We use $\cos^{4}$-like pulses which are close to Gaussian but restricted in time as real laser pulses,
\begin{equation}
f(\eta)=\left\{
\begin{array}
[c]{ll}%
\cos^{4}(\pi{}\eta), & |\eta{}|<1/2\\
0, & |\eta{}|>1/2
\end{array}
\right..\label{eq:fcos}%
\end{equation}
The function $f(\eta)$ is close to the Gaussian
distribution $f_{G}(\eta)=\exp\left(  -2\pi^{2}\eta^{2}\right)$ in the interval where $f_{G}(\eta)$ is not very small.
The area of the pulse with the envelope described by function (\ref{eq:fcos}) equals $\frac{3}{8}\Omega_{0}\tau$, 
that is approximately
0.94 times the area of the corresponding Gaussian pulse. The characteristic width of the latter 
is $\tau_{G}\approx0.225\tau$. 
More close adjustment of  $\cos^{n}$-like pulse to the Gaussian pulse is possible:
the function $\cos^{n}(\pi t/\tau)$ tends to $\exp(-t^{2}/\tau_{G}^{2})$ with $\tau_{g}=\tau\sqrt{2}(\pi\sqrt{n})^{-1}$
for large even $n$  within the interval $\left\vert t\right\vert <\tau/2$~\cite{Rom06}.

\section{The wave function calculation}
\label{sec-MCWF}

We describe the atomic state by the wave function which is constructed by the Monte-Carlo wave function (MCWF)
method \cite{Mol93}. 
After averaging  over the ensemble of atoms or molecules, this approach becomes equivalent
to the solution of the density matrix equation. At the same time,
in contrast to the latter, it allows one to give an illustrative interpretation for  the separate atom's or molecule's motion.

The wave function obeys the Schr\"odinger equation 
\begin{equation}
i\hbar\frac{\mathrm{d}}{\mathrm{d}t}|\psi\rangle=H|\psi\rangle,\label{eq:Sch}
\end{equation}
where the Hamiltonian 
\begin{eqnarray}
H&=& \hbar\varpi_{1}|e\rangle\langle{}e|+\hbar\Delta|g_{2}\rangle\langle{}g_{2}|
 -\bm{\mathrm{d}}_{g_{1}e}|g_{1}\rangle\langle{}e|\bm{\mathrm{E}}(t)                     \nonumber\\
&&-\bm{\mathrm{d}}_{eg_{1}}|e\rangle\langle{}g_{1}|\bm{\mathrm{E}}(t)-
\bm{\mathrm{d}}_{g_{2}e}|g_{2}\rangle\langle{}e|\bm{\mathrm{E}}(t) \nonumber\\
&&-\bm{\mathrm{d}}_{eg_{2}}|e\rangle\langle{}g_{2}|\bm{\mathrm{E}}(t)-\frac{i\hbar}{2}\left(\gamma_{1}+\gamma_{2}\right)|e\rangle\langle{}e| ,
\label{eq:Ham}
\end{eqnarray}
which is used for the construction of the wave function by MCWF method, 
differs in the relaxation term from the Hamiltonian which is used in the density matrix equation. 

Hamiltonian (\ref{eq:Ham}) is non-Hermitian, hence the absolute value of
the wave function determined from the Schr\"{o}dinger equation~(\ref{eq:Sch})  changes with
time. In the MCWF method, normalization of the wave function should be carried out after every small time step. 
Besides that, the condition of a quantum jump within each time interval has to be checked \cite{Mol93}.

We use the first order method for calculation of Monte-Carlo wave function~\cite{Mol93}.
More precise the second and the fourth order methods are described in~\cite{Ste95}.

Let the wave function $|\psi(t)\rangle$ is normalized to unity at the time moment $t$. 
After a small time step $\Delta t$ the wave function $|\psi(t)\rangle$ is transformed into
\begin{equation}
|\psi^{(1)}(t+\Delta{}t)\rangle=\left(1-\frac{i\Delta{}t}{\hbar}{{H}}\right)|\psi(t)\rangle
\label{eq:phiI}
\end{equation}
according to  Schr\"{o}dinger equation (\ref{eq:Sch}).
The squared norm of the wave function equals
\begin{equation}
\langle\psi^{(1)}(t+\Delta{}t)|\psi^{(1)}(t+\Delta{}t)\rangle=1-\Delta{}P,
\label{eq:phiIN}
\end{equation} 
where
\begin{eqnarray}
\Delta{}P&=&\frac{i\Delta{}t}{\hbar}\langle\psi(t)|H-H^{+}|\psi(t)\rangle\nonumber\\
&=&\left(\gamma_{1}+\gamma_{2}\right)\langle\psi(t)|e\rangle\langle{}e|\psi(t)\rangle\Delta{}t.
\label{eq:dPpsi}
\end{eqnarray}

Now we take into account a possibility of quantum jump. 
If the value of the random variable $\epsilon$, which is uniformly distributed between zero and unity, is larger than 
$\Delta{}P$ (it is true in the most cases, as far as $\Delta{}P\ll1$), there is no  jump. Then the wave function at the time moment $t+\Delta t$ equals
\begin{equation}
|\psi(t+\Delta{}t)\rangle=|\psi^{(1)}(t+\Delta{}t)\rangle/\sqrt{1-\Delta{}P}, \quad \Delta{}P<\epsilon.
\end{equation}
In the opposite case, $\epsilon<\Delta P$, the jump occurs, and the wave functions becomes
\begin{equation}
|\psi(t+\Delta{}t)\rangle=|g_{1}\rangle
\label{eq:g1}
\end{equation}
with probability $p_{1}=\gamma_{1}/(\gamma_{1}+\gamma_{2})$, or
\begin{equation}
|\psi(t+\Delta{}t)\rangle=|g_{2}\rangle
\label{eq:g2}
\end{equation}
with probability $p_{2}=\gamma_{2}/(\gamma_{1}+\gamma_{2})$.  If the value of  the
random variable $\epsilon$, uniformly distributed between 0 and 1,  is less then $p_{1}$, the wave function is (\ref{eq:g1}), otherwise it is (\ref{eq:g2}).

It is convenient to separate the rapid component, varying with the frequency $\varpi_{1}$,  in the wave function.  
For this purpose we seek for the solution of~(\ref{eq:Sch}) in the form 
\begin{equation}
|\psi\rangle=C_{g_{1}}|g_{1}\rangle+C_{{2}}e^{-i\Delta{}t}|g_{2}\rangle+C_{e}e^{-i\varpi_{1}t}|e\rangle.
\label{eq:psi}
\end{equation}
After applying rotating wave approximation~\cite{Sho90} to the Schr\"odinger equation we find, assuming $\Delta\ll\varpi_{1}$, 
the equations  for probability amplitudes 
\begin{eqnarray}
\frac{d}{dt}C_{g_{1}}&=&-\frac{i}{2}\Omega_{1}e^{-ikz-i\delta_{1}t}\sum\limits_{m}e^{i\varphi_{11,m}}f(\eta_{1,m})C_{e}\nonumber{}\\
&&-\frac{i}{2}\Omega_{1}e^{ikz-i\delta_{1}t}\sum\limits_{m}e^{i\varphi_{12,m}}f(\eta_{2,m})C_{e}\label{eq:cgi}\\
\frac{d}{dt}C_{g_{2}}&=&-\frac{i}{2}\Omega_{2}e^{-ikz-i\delta_{2}t}\sum\limits_{m}e^{i\varphi_{21,m}}f(\eta_{1,m})C_{e}\nonumber\\
&&-\frac{i}{2}\Omega_{2}e^{ikz-i\delta_{2}t}\sum\limits_{m}e^{i\varphi_{22,m}}f(\eta_{2,m})C_{e},\label{eq:cgii}\\
\frac{d}{dt}C_{{e}}&=&-\frac{i}{2}\Omega_{1}e^{ikz+i\delta_{1}t}\sum\limits_{m}e^{-i\varphi_{11,m}}f(\eta_{1,m})C_{g_{1}}\nonumber\\
&&-\frac{i}{2}\Omega_{1}e^{-ikz+i\delta_{1}t}\sum\limits_{m}e^{-i\varphi_{12,m}}f(\eta_{2,m})C_{g_{1}}\nonumber\\
&&-\frac{i}{2}\Omega_{2}e^{ikz+i\delta_{2}t}\sum\limits_{m}e^{-i\varphi_{21,m}}f(\eta_{1,m})C_{g_{2}}\nonumber\\
&&-\frac{i}{2}\Omega_{2}e^{-ikz+i\delta_{2}t}\sum\limits_{m}e^{-i\varphi_{22,m}}f(\eta_{2,m})C_{g_{2}}\nonumber\\
&&-\frac{\gamma_{1}+\gamma_{2}}{2}C_{e},
\label{eq:ce}
\end{eqnarray} 
which are to be solved numerically simultaneously with the quantum jump testing.

Most time (during the time interval between the light pulses) the field does not influence  the atom or the molecule.
In this case the analytical solution of the Eqs. ~(\ref{eq:cgi})--(\ref{eq:ce}) is possible.  
Let the initial atom's or molecule's state is
\begin{equation}
|\psi(0)\rangle=C_{g_{1}}(0)|g_{1}\rangle+C_{g_{2}}(0)|g_{2}\rangle+C_{e}(0)|e\rangle.
\label{eq:psi0}
\end{equation}   
If no quantum jump occurs within the time interval $[0,t]$,  we find from the Eqs.\ (\ref{eq:cgi})--(\ref{eq:ce}) the normalized
wave function
\begin{eqnarray}
|\psi(t)\rangle&=&C_{g_{1}}(t)|g_{1}\rangle+C_{g_{2}}(t)e^{-i\Delta{}t}|g_{2}\rangle\nonumber\\
&&+C_{e}(t)e^{-i\varpi_{1}t}|e\rangle,
\label{eq:Sch-sol-psi}
\end{eqnarray}
where
\begin{eqnarray}
C_{g_{1}}(t)&=&{C_{g_{1}}(0)}/{D},\label{eq:Sch-gi}\\
C_{g_{2}}(t)&=&{C_{g_{2}}(0)}/{{D}},\label{eq:Sch-gii}\\
C_{e}(t)&=&{C_{e}(0)e^{{-\frac{1}{2}\left(\gamma_{1}+\gamma_{2}\right)t}}}/{D}
\label{eq:Sch-e}
\end{eqnarray}
and
\begin{equation}
D= \sqrt{|C_{g_{1}}(0)|^{2}+|C_{g_{2}}(0)|^{2}+|C_{e}(0)|^{2}e^{-(\gamma_{1}+\gamma_{2})t}}.
\label{eq:D}
\end{equation}

The probability of the absence of a quantum jump within the time interval $[0, t]$ is~\cite{Mol93}%
\begin{equation}
P(t)=|C_{1}(0)|^{2}+|C_{2}(0)|^{2}+|C_{e}(0)|^{2}e^{-(\gamma_{1}+\gamma_{2})t}.
\label{eq:P}
\end{equation}
The expression~(\ref{eq:P})  is consistent with the probability $|C_{1}(0)|^{2}+|C_{2}(0)|^{2}$ of no quantum
jump for $t\to\infty$ and the exponential decay of the excited state in the ensemble of atoms or molecules.

So, in sum, in the absence of laser radiation within the time interval $[0, t]$ the atom or molecule
is described by the state~(\ref{eq:Sch-sol-psi}) at the time instant $t$ with the probability~(\ref{eq:P}).
The other possible states are  
\begin{equation}
|\psi(t)\rangle=|g_{1}\rangle
\label{eq:psi-1-P1}
\end{equation}
with the probability $\gamma_{1}\left[1-P(t)\right]/\left(\gamma_{1}+\gamma_{2}\right)$ and 
\begin{equation}
|\psi(t)\rangle=|g_{2}\rangle.
\label{eq:psi-1-P2}
\end{equation}
with the probability $\gamma_{2}\left[1-P(t)\right]/\left(\gamma_{1}+\gamma_{2}\right)$.       

\section{Atom's and molecule's motion}
\label{sec-mech}

Cooling of atoms in one-dimensional molasses was successfully simulated by MCWF method in~\cite{Mol93}.
In this case only the atomic momentum distribution function matters. 
Analyzing possible simultaneous cooling and trapping of 
atoms or molecules in the considering trap, 
we need both the spatial and momentum distribution functions. 
Quantum-mechanical calculation of the atomic 
motion in the trap should start from the wave package with spatial width much less then the laser radiation wavelength.
As consequence, a lot of momentum states of the atom both in the ground an excited state 
are involved in the calculation. 

The computation time can be substantially reduced for the case of weak 
laser fields, when the momentum diffusion of the atoms could be treated as 
caused by counter-propagating laser pulses independently. In this case we consider the atom's motion in the framework 
of classical mechanics and need to know the light pressure force, which the atoms undergo. 
This force can be calculated from the density matrix and the electric field of the pulses~\cite{Min87,Met99},
\begin{equation}
F=\sum\limits_{j=1}^{2}(\varrho_{g_{j}e}\bm{\mathrm{d}}_{eg_{j}}+\varrho_{eg_{j}}\bm{\mathrm{d}}_{g_{j}e})\frac{\partial \bm{\mathrm{E}}}{\partial z},
\label{eq:F}
\end{equation}
where the density matrix elements are expressed in terms of $C_{g_{1}}$, $C_{g_{2}}$ and
$C_{e}$ as follows:%
\begin{eqnarray}
\varrho_{g_{j}g_{j}}&=&|C_{j}|^2,\quad{} j=1,2, \\
\varrho_{ee}&=&|C_{e}|^2, \\
\varrho_{eg_{1}}&=&C_{1}^{*}C_{e}e^{-i\varpi_{1}{}t}, \\
\varrho_{g_{1}\!e}&=&C_{1}C_{e}^{*}e^{i\varpi_{1}{}t}, \\
\varrho_{eg_{2}}&=&C_{2}^{*}C_{e}e^{-i\varpi_{2}{}t}, \\
\varrho_{g_{2}e}&=&C_{2}C_{e}^{*}e^{i\varpi_{2}{}t}.
\end{eqnarray}
We assume that the pulse duration considerably exceeds the inverse carrier frequency, 
$\omega_{1}\tau\gg1$, $\omega_{2}\tau\gg1$, 
therefore we neglect the derivative of the pulse's envelope in calculation of the time derivative of the field strength, as far as
$|\frac{\partial f(\eta_{j,m})}{\partial z}|\ll{}k_{j}f(\eta_{j,m})$ ($j=1,2$).

After averaging over the period of oscillations with the frequency $\omega_{1}$, 
the expression (\ref{eq:F}) in the field (\ref{eq:E}) gives
\begin{eqnarray}
{{{F}}}&=&\left[\hbar{}k_{1}\Omega_{1}\mathop{\mathrm{Im}}C_{1}C_{e}^{*}e^{i\delta_{1}t+ik_{1}z}\sum\limits_{m}e^{-i\varphi_{11,m}}f(\eta_{1,m})\right.\nonumber\\
&&+\hbar{}k_{2}\Omega_{2}\mathop{\mathrm{Im}}C_{2}C_{e}^{*}e^{i\delta_{2}t+ik_{2}z}\sum\limits_{m}e^{-i\varphi_{21,m}}f(\eta_{1,m})\nonumber\\
&&-\hbar{}k_{1}\Omega_{1}\mathop{\mathrm{Im}}C_{1}C_{e}^{*}e^{i\delta_{1}t-ik_{1}z}\sum\limits_{m}e^{-i\varphi_{12,m}}f(\eta_{2,m})\nonumber\\
&&\left.-\hbar{}k_{2}\Omega_{2}\mathop{\mathrm{Im}}C_{2}C_{e}^{*}e^{i\delta_{2}t-ik_{2}z}\sum\limits_{m}e^{-i\varphi_{22,m}}f(\eta_{2,m})\right]\nonumber\\
&&\times\left[|C_{1}|^{2}+|C_{2}|^{2}+|C_{e}|^{2}\right]^{-1}.
\label{eq:FF}
\end{eqnarray}   

The dependences of the atom's coordinate $z$ on time we find from the Newton's equation
\begin{equation}
\ddot{z}=F/M,
\label{eq:ddz}
\end{equation}
where  $M$ is the atom's mass. We consider the case $\left|\varpi_{1}-\varpi_{2}\right|\ll\varpi_{1}$,
and assume $k_{1}=k_{2}$ in~(\ref{eq:FF}).  

The Eq.~\eqref{eq:ddz} does not take into account 
the momentum change due to spontaneous emission of photons.
Every event of spontaneous emission of a photon change 
the atom's or the molecule's velocity by $\hbar{}\bm{k}/M$ in 
the random direction with the probability $1-P(t)$, where $P$ is determined by~(\ref{eq:P}). Besides that,
the velocity also changes due to fluctuations of absorption and stimulated emission of photons. 

\section{Numerical calculation routine}
\label{sec-num}

To simulate the atom's or molecule's motion, we simultaneously solve the 
Eqs.~\eqref{eq:cgi}--\eqref{eq:ce} and~(\ref{eq:ddz}), where the light pressure force we 
find from~(\ref{eq:FF}). Besides that, we take into account
both the atomic momentum's change
due to the spontaneous emission of photons and fluctuation of stimulated (absorption and emission) processes.  
In our model calculations we postulate that spontaneous emission occurs with equal probability 
in two directions along the light beam, so the atom's or molecule's momentum changes by $ \pm \hbar k$.  
This assumption in analyses of Doppler cooling leads to minimum temperature~\cite{Ada97}
\begin{equation}
T_{\min}= \hbar\gamma/2k_{B},
\label{eq:Doppler}
\end{equation} 
where $ k_{B} $ is the Boltzmann constant, $ \gamma $ is the rate of 
the spontaneous emission of radiation by the excited atom. 

The light pressure force~\eqref{eq:F} gives correct value for the ensemble averaged force, but 
the momentum diffusion phenomenon is not correctly taken into account. To analyze the motion 
of atoms or molecules in the trap we need to add stochastic change of the momentum, zero in average,
that gives correct momentum diffusion coefficient. We analyze the low intensity case,
when the population of the excited state is small and the light pressure force and 
the momentum diffusion approximately equal to the sum of the corresponding values for each of the 
counter-propagating traveling waves. Here we describe the momentum diffusion of atoms in the field of 
one traveling wave following~\cite{Min87}. 

Let the momentum of an atom at the time instant $ t $ is $ \mathbf{p}_{0} $. Then at the time instant $ t+\Delta t $
the momentum is
\begin{equation}
\mathbf{p}=\mathbf{p}_{0}+\hbar{}\mathbf{k}(N_{+}-N_{-})-\sum_{s}\hbar{}\mathbf{k}_{s}.
\label{eq:p}
\end{equation}
Here the second term gives the change of momentum due absorption and stimulated emission,
when the photons with the wave vectors $ \mathbf{k} $ (directed along $ z $-axis) are absorbed and emitted. 
The quantities  $N_{+}$ and $N_{-}$ are the numbers of the absorbed and emitted photons. 
The third term in~\eqref{eq:p} is responsible for the momentum change due to the spontaneous emission of
the photons with the wave vectors $ \mathbf{k}_{s} $.

The ensemble average of the momentum~\eqref{eq:p} is
\begin{equation}
\langle\mathbf{p}\rangle=\langle\mathbf{p}_{0}\rangle+\hbar{}\mathbf{k}(\langle N_{+}\rangle-\langle N_{-}\rangle),
\label{eq:pav}
\end{equation}
where $ \langle\mathbf{p}_{0}\rangle $ is the initial average momentum, $ \langle N_{+}\rangle $ is the average number 
of the absorbed photons, $ \langle N_{-}\rangle $ is the average number of the photon emitted by atoms 
in the process of stimulated emission. The photons emitted in the process of spontaneous emission 
does not change the average momentum,
\begin{equation}
\Bigl\langle\sum_{s}^{}\mathbf{k}_{s}\Bigr\rangle=0.
\label{eq:ksav}
\end{equation}
The difference of \eqref{eq:p} and \eqref{eq:pav} gives the momentum fluctuation,
\begin{equation}
\Delta \mathbf{p} =\mathbf{p}-\langle\mathbf{p}\rangle=
(\mathbf{p}_{}-\langle\mathbf{p}_{0}\rangle)+\hbar\mathbf{k}\Delta N_{i}-\sum_{s}^{}\hbar\mathbf{k}_{s},
\label{eq:Dp} 
\end{equation}
where $\Delta N_{i} = N_{i}-\langle N_{i} \rangle$ is the variation of the difference $ N_{i}=N_{+}-N_{-} $ 
from the corresponding ensemble average value. 

The average square of the momentum fluctuations along $ z $-axis is 
\begin{equation}
\langle\Delta {p}_{z}^{2}\rangle =\langle\Delta {p}_{0z}^{2}\rangle+
\hbar^{2}{k}^{2}\langle(\Delta N_{i})^{2}\rangle+\hbar^{2}{k}^{2}\langle\cos^{2}\theta\rangle\langle N_{s}\rangle.
\label{eq:DpDp} 
\end{equation}
Here $ \theta $ is the angle between the direction of the photon's spontaneous emission and $ z $-axis,
$ \langle N_{s}\rangle $ is the average number of the spontaneously emitted photons.
The first term in the r.h.s. of~\eqref{eq:DpDp} gives the initial momentum spreading, the second term
is due to stimulated processes (absorption and emission), the third term is due to spontaneous emission. 

Let's find $ \langle(\Delta N_{i})^{2}\rangle $ assuming the Poisson photons statistics. In this case
\begin{equation}
\langle(\Delta N_{i})^{2}\rangle=\langle N_{i}\rangle.
\label{eq:Poisson}
\end{equation}
Noting that $ \langle N_{i}\rangle=\langle N_{s}\rangle $, we finally find
\begin{equation}
\langle\Delta {p}_{z}^{2}\rangle =\langle\Delta {p}_{0z}^{2}\rangle+
\hbar^{2}{k}^{2}\langle N_{s}\rangle+\hbar^{2}{k}^{2}\langle\cos^{2}\theta\rangle\langle N_{s}\rangle.
\label{eq:DpDpf} 
\end{equation}
This equation shows the way for numerical modeling of momentum diffusion process in the field of traveling 
wave. Each  random momentum change due to spontaneous emission is accompanied by stimulated process,
in which the momentum of the atom is changed by $ \pm\hbar k $.

Now consider the case of counter-propagating laser pulses.  
When counter-propagating laser pulses are weak, spontaneous emission follows each absorbed photon, 
so the fluctuation events of the atomic or molecular velocity change due to light induced processes
occur as frequently as events of spontaneous emission. This point is the background of 
our computer simulation of atoms and molecules movement in the field of laser radiation. 

In our calculation we assume the model of $ \pm \hbar k $ change of the momentum due to spontaneous emission 
($ \theta $ equals $ 0 $ or $ \pi $ with equal probability).
We use different approaches to solving these Eqs.~\eqref{eq:cgi}--\eqref{eq:ce} and~(\ref{eq:ddz}), (\ref{eq:FF})
during the atom's interaction with the pulses and free evolution of the atom.
In the first case the solution to the set of equations is found by using Runge-Kutta fourth order 
method with fixed step size.
After every step we check if a quantum jump occurs and normalize the wave function. If a jump occurs, 
the atom's velocity changes by 
$\Delta{}v=\hbar{}k(\epsilon_1-0.5)|/(M|\epsilon_1-0.5|)+\hbar{}k(\epsilon_2-0.5)|/(M|\epsilon_2-0.5|)$, 
where $\epsilon_{1,2}$ are random numbers with a uniform distribution in the interval $[0,1]$. 
In the second case, when the field does not act on the atom,
we do not need to divide the considered time interval by small subintervals and check if the quantum jump occurs in every subinterval. 
Knowing the probability (\ref{eq:P}) of the absence of a quantum jump 
within the time interval $[0, t]$,
we simulate the time moment of the quantum jump. The scheme of calculation is following. We compare the value of 
the uniformly distributed in the interval $[0,1]$ random variable $\epsilon$ with $|C_{1}(0)|^{2}+|C_{2}(0)|^{2}$ 
at the beginning of the time interval. A jump occurs if $\epsilon>|C_{1}(0)|^{2}+|C_{2}(0)|^{2}$, 
and does not otherwise. In the latter case 
the wave function is described by Eqs.~(\ref{eq:psi}),  (\ref{eq:Sch-gi})--(\ref{eq:Sch-e}).
If a jump occurs, we simulate the time moment of the quantum jump. 
We take a random $ \epsilon $, which is uniformly distributed in the interval $[0,1]$. 
For the exponential distribution of probability, $P_{e}=e^{-(\gamma_{1}+\gamma_{2})t}$, 
the quantity $t_{jump}=-\left(  \ln\epsilon\right)(\gamma_{1}+\gamma_{2})^{-1}$ simulates the time moment 
when the jump occurs~\cite{Sob73}.
If $t_{jump}$ exceeds the time interval between the laser pulses, we calculate the probability 
amplitudes (\ref{eq:Sch-gi})--(\ref{eq:Sch-e}) 
at the beginning of the next pulse, otherwise we calculate the atom's velocity 
change $\Delta{}v=\hbar{}k(\epsilon_1-0.5)|/(M|\epsilon_1-0.5|)+\hbar{}k(\epsilon_2-0.5)|/(M|\epsilon_2-0.5|)$
at $t_{jump}$ using random numbers $\epsilon_{1,2}$ with a uniform distribution in the interval $[0,1]$.
The atom's or molecule's state is~(\ref{eq:psi-1-P1}), with the probability  
$\gamma_{1}\left[1-P(t)\right]/\left(\gamma_{1}+\gamma_{2}\right)$, or~(\ref{eq:psi-1-P2}), with the probability  
$\gamma_{2}\left[1-P(t)\right]/\left(\gamma_{1}+\gamma_{2}\right)$. To choose between these states, we compare
$\gamma_{1}/(\gamma_{1}+\gamma_{2})$ with a new random value $\epsilon$.
When $\epsilon<\gamma_{1}/(\gamma_{1}+\gamma_{2})$, the state
of the atom or the molecule is described by~(\ref{eq:psi-1-P1}), 
otherwise by wave function~(\ref{eq:psi-1-P2}).

The described approach substantially reduces the calculation time in comparison with Runge-Kutta method 
during whole time of the atom's or the molecule's motion.

To estimate the temperature of the captured atoms or molecules, we average the velocity and the 
squared velocity over the ensemble of particles.
                                                            
\section{Results and discussion}
\label{sec-Res}

In this section we describe the results of the numerical simulation of atoms and molecules motion 
in the trap formed by the trains of counter-propagating light pulses. 
In contrast to the results of~\cite{Rom13,Rom14}, where the evolution of two-level atoms  was
investigated, here we also study statistical characteristics of the atomic and molecular ensembles. 

We analyze the simplest models of the atom-field and the molecule-field interaction. 
It is well known that the cycling atom-field interaction 
can be realized between two states of some atoms~\cite{Met99}. 
As an example of such interaction, we chose transition $3^{2}S_{1/2}{-}3^{2}P_{3/2}$ 
in the sodium atom. The simplest molecule-field interaction model include three levels. 
The transitions coupling the state $A^{2}\Pi_{1/2}(v'=0,J'=1/2)$ with 
the states $X^{2}\Sigma_{1/2}^{+}(v=0,N=1)$ and $X^{2}\Sigma_{1/2}^{+}(v=1,N=1)$ of $\mathrm{SrF}$ 
form  the almost close three-level 
$\Lambda$-scheme~\cite{Bar12}. The spontaneous emission from the upper state leads 
the molecule to the lower states with the probability 0.9996. 
Including the state $X^{2}\Sigma_{1/2}^{+}(v=2,N=1)$ into the considered model gives 
the probability of the spontaneous transition to the three lower states more then
$0.9999$, but we do not add this state, possibly sacrificing the simulation accuracy 
for the sake of greater physical clarity. Anyway, an additional light fields can return the 
molecules which is lost from the scheme due to the spontaneous emission, 
as it was realized in experiment~\cite{Bar12}. 

\subsection{Two-level model}
Nowadays the investigation of simultaneous trapping and cooling of atoms by the counter-propagating laser pulses
are presented in two papers,~\cite{Rom13,Rom14}, for the two-level model of the atom-field interaction.  
The authors of the first paper studied the momentum diffusion of the two-level atoms 
in an optical trap formed by sequences of the counter-propagating light pulses trains and discovered that proper detuning 
of carrier frequency of laser pulses from the resonance with the atom's transition frequency leads 
to cooling of the atomic ensemble. 
The other sign of the detuning, as well as the resonant interaction of the field with atoms, 
leads to ``heating'' of the atomic ensemble. 
The conclusions of~\cite{Rom13} are based on the computer simulation of the motion of an atom
in the trap for hypothetical atomic and atom-field interaction parameters. 
In~\cite{Rom14} the motion of $\mathrm{^{23}Na}$ atom in the trap was analyzed. 
Here we take the next step in the pulse trap investigation, 
which includes the simulation of the atomic ensemble characteristics.

Possible cooling of atoms in the trap can be easily explained for weak pulses, $ \vartheta\ll\gamma T $,
where $ \vartheta\equiv\vartheta_{1} $ is the pulse area, $ \gamma\equiv \gamma_{1} $.
In this case the atom mostly interacts with the spectral component of the pulses trains which is closest to 
the transition frequency. Let the carrier frequency of the pulses 
is tuned below the transition frequency in the atom.
Then the atoms, due to Doppler effect, always absorb more photons 
from the laser beam opposite to their direction of motion. 
As a result, ``a friction force'' arises and cools 
the atoms down to the Doppler cooling limit~\eqref{eq:Doppler}.
This limit is caused by the competition between the cooling 
due to the friction force and heating due to the momentum diffusion. 
For large pulse areas the detuning  of the carrier frequency
from the resonance with the transition frequency in the atom, needed for atoms cooling, change sign~\cite{Rom14}.
 
\begin{figure}[ht]
\begin{center}
\includegraphics[width=87mm]{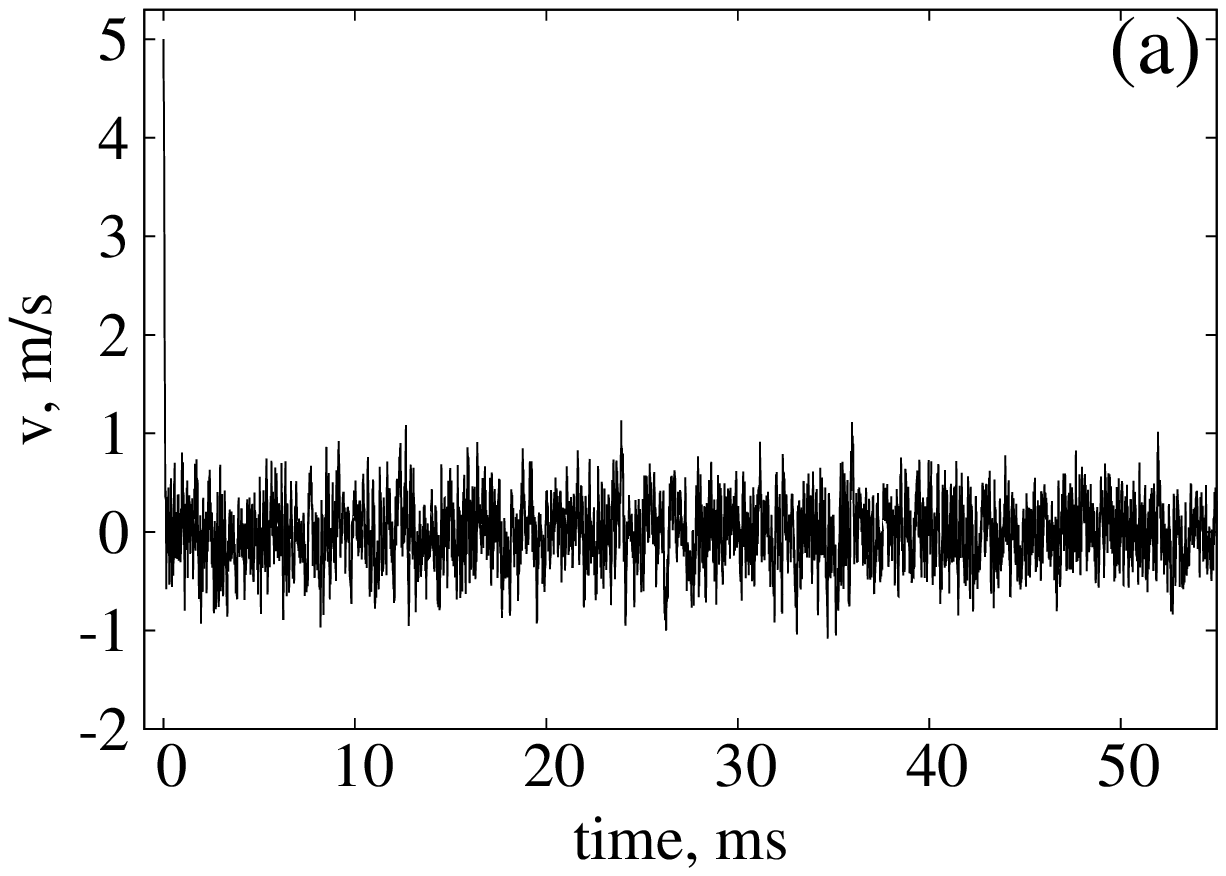}
\bigskip

\includegraphics[width=87mm]{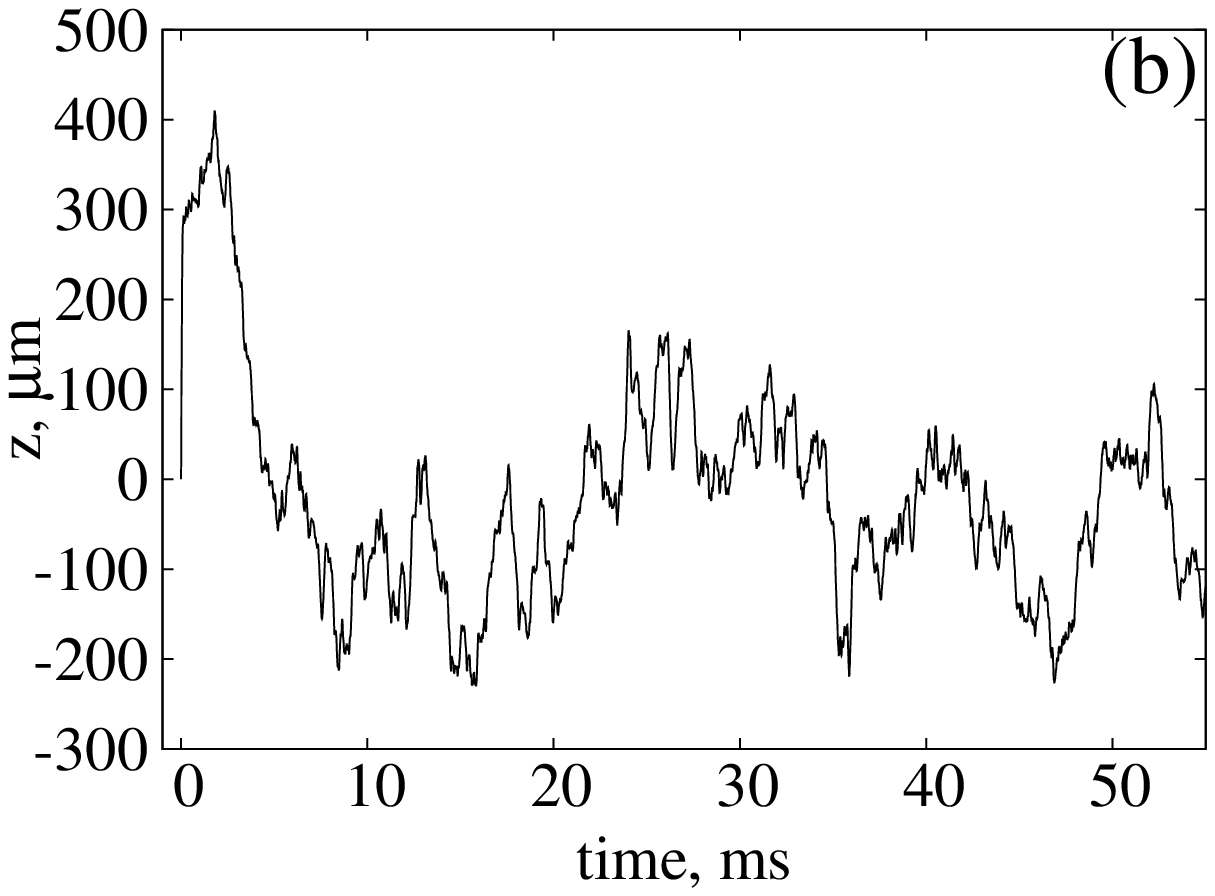}

\caption{An example of the sodium atom's motion in the field of the counter-propagating sequences of light pulses: 
(a) $ v(t) $ and  (b) $ z(t) $. Parameters: $ \tau=1 $~ps, $ T=10 $~ns, $ \vartheta=0.05\pi $,
$ \gamma=2\pi\times10 $~MHz, $ \delta=2\pi\times5 $~MHz, $ \lambda=589 $~nm, $ M=23 $~Da. The atom starts
at the center of the trap with initial velocity $ v_{0}=5 $~m/s.}%
\label{figure:3}
\end{center}
\end{figure}

We consider an optical trap which extends from $z=-100$~mm to $z=100$~mm relative to the point, 
where counter-propagating light pulses ``collide'', and trace the motion of an atom until it moves inside the trap.
Level $g_{2}$ is not taken into account in the simulation of
$ ^{23}\mathrm{Na}$ motion in the trap. Besides that,
we suppose $\varphi_{11,m}=\varphi_{12,m}=0$ in~(\ref{eq:cgi}), (\ref{eq:ce}). Figure~\ref{figure:3} 
shows an example of the atom's motion in the field of the counter-propagating sequences 
of 1-ps light pulses with repetition frequency 100~MHz.
Very quickly (0.14~ms after the beginning of the interaction with the field) the atom slows down to zero velocity
and then its velocity fluctuates in the region $ \pm 1$~m/s. The atom returns to the center of the trap
approximately after 4.7~ms and then fluctuates in the region $ \pm 0.25$~mm. The velocity capture range of the trap
for the parameters specified in Fig.~\ref{figure:3} extends at least from $v=-20 $~m/s to $v= +20 $~m/s 
(temperature of atoms about 1~K).

\begin{figure}[ht]
\begin{center}
\includegraphics[width=87mm]{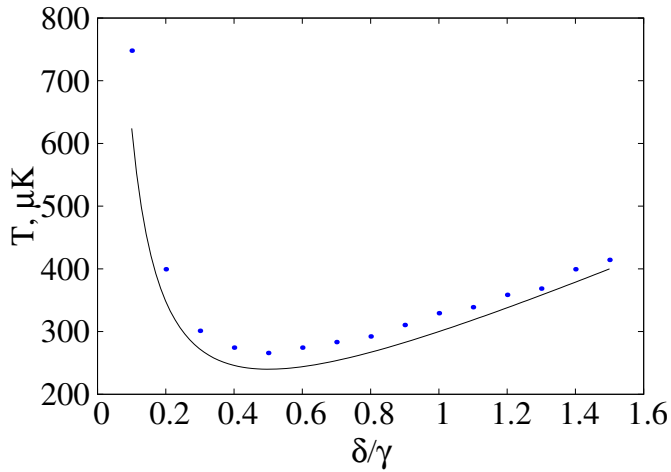}

\caption{The dependence of the temperature of 1000 sodium atom's in the trap formed by the counter-propagating 
sequences of light pulses on the pulse's carrier frequency detuning on the atomic transition frequency.
Parameters: $ \tau=1 $~ps, $ T=10 $~ns, $ \vartheta=0.05\pi $,
$ \gamma=2\pi\times10 $~MHz, $ \lambda=589 $~nm, $ M=23 $~Da.}%
\label{figure:4}
\end{center}
\end{figure}

To estimate the measure of cooling in the trap, we introduce the ``temperature'' 
of the atomic ensemble by the expression
\begin{equation}
T_{a}=\frac{m\langle{}v^{2}\rangle}{k_{B}},
\label{eq:R-Ta}
\end{equation}
where $k_{B}$ is the Boltzmann constant. 
The value of $T_{a}$ coincides with the real temperature of the ensemble
in the case of Maxwell velocity distribution.
We expect that cooling process in the trap is close to the Doppler cooling~\cite{Ada97,Met99}, anyway
for the case of weak field. This expectation is confirmed by comparison of smooth curve and dots
in the Fig.\ref{figure:4}, where the dots were calculated from Eq.~\eqref{eq:R-Ta} with averaging over 
1000 sodium atoms
and smooth curve represents the dependence of the atoms temperature on the detuning
of the frequency of the weak monochromatic standing wave from the atomic transition frequency~\cite{Ada97}
\begin{equation}
T_{sw}=\frac{1}{2}T_{\min}\left (\frac{2\delta}{\gamma}+\frac{\gamma}{2\delta}\right ).
\label{eq:Tsw}
\end{equation}
For sodium atoms $ T_{\min}=240\,\mu$K~\cite{Met99}.
The temperature is minimal, as in the case of the standing wave, for $ \delta=\gamma/2 $.
The difference between the curve and the dots, according to our calculations, becomes less for smaller pulse's areas.

\begin{figure}[ht]
\begin{center}
\includegraphics[width=87mm]{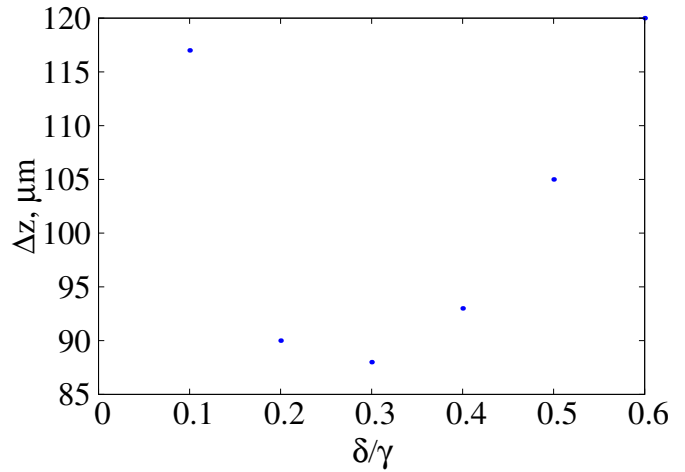}

\caption{The dependence of the spatial capture range of the trap
$ \Delta z= \sqrt{\left\langle z^{2}\right\rangle -\left\langle z\right\rangle^{2}}$
on the frequency detuning $ \delta $. The parameters are the same as in Fig.~\ref{figure:4}.
}%
\label{figure:5}
\end{center}
\end{figure}

\begin{figure}[ht]
\begin{center}
\includegraphics[width=87mm]{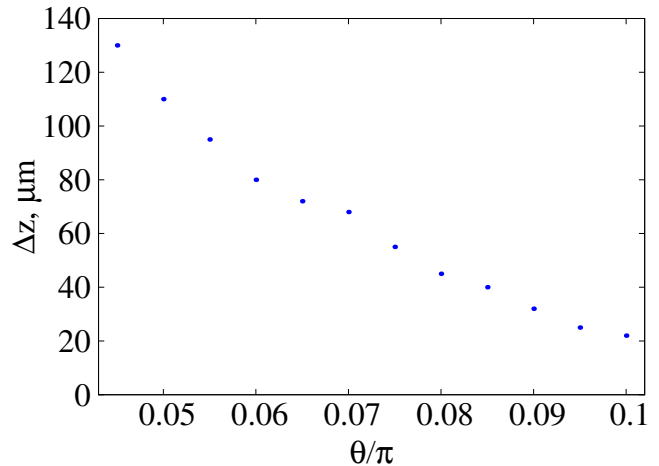}

\caption{The dependence of the spatial capture range of the trap
$ \Delta z= \sqrt{\left\langle z^{2}\right\rangle -\left\langle z\right\rangle^{2}}$
on the pulse area for $ \delta =0.5\gamma$. The other parameters are the same as in Fig.~\ref{figure:4}.
}%
\label{figure:6}
\end{center}
\end{figure}

The spatial capture range of the trap depends on $ \delta $ and $ \vartheta$. 
The first dependence is depicted in Fig.~\ref{figure:5}.
The minimal capture range does not coincide with the minimal temperature; it reaches
approximately for $ \delta=0.3\gamma $. For the parameters used in modeling this dependence,
the atoms are localized in the region of the pulses' overlapping. This region became narrower
when pulse area increases (see Fig.~\ref{figure:6}).
                                                                                                                                                                                                                                       
\subsection{Three-level model}

We simulate the dynamics of the three level system for the parameters, that are close 
to the interaction of $\mathrm{SrF}$ with CW laser radiation~\cite{Bar12}. Our consideration 
neglects the probability 0.0004 of the molecule to leave the $ \Lambda $-scheme in the process
of spontaneous emission from the excited level (see Fig.~\ref{figure:2}).
The spontaneous emission rate from the excited state is
$ \gamma=\gamma_{1}+\gamma_{2} =2\pi\times7$~MHz, branching ratio $ \gamma_{2}/\gamma_{1}=0.02$.
Considering the equal energy of the pulses, we came to conclusion that 
$ \vartheta_{2}/ \vartheta_{1}=d_{eg_2}/d_{eg_1}=\sqrt{\gamma_{2}/\gamma_{1}}=0.14$.
The wavelengths of the transitions are  $ \lambda_{1}=663.3 $~nm ($ e\Leftrightarrow g_1$),
$ \lambda_{2}=686.0 $~nm ($ e\Leftrightarrow g_2$). In calculation of the photon momentum for each transition we
neglect the difference between $ \lambda_{1} $ and $ \lambda_{2} $.
Repetition period of the pulses is chosen $ T=23.8 $~ns. It corresponds to the period of frequency modulation 
of laser radiation in the experiment~\cite{Bar12}, that provides the excitation of all superfine sublevels 
of the ground states. 
Detunings $ \delta_{1} $ and $ \delta_{2} $ should not correspond to the two-photon resonance condition 
$ \delta_{2}-\delta_{1}=2\pi n/T$, where $ n $ is integer, to avoid the coherent population trapping, 
otherwise the population of the excited state
becomes zero and the light pressure force vanishes~\cite{Ili12,Ili12-a}.
As in the case of the two-level model, the pulse  duration is $\tau=1 $~ps. 

\begin{figure}[ht]
\begin{center}
\includegraphics[width=87mm]{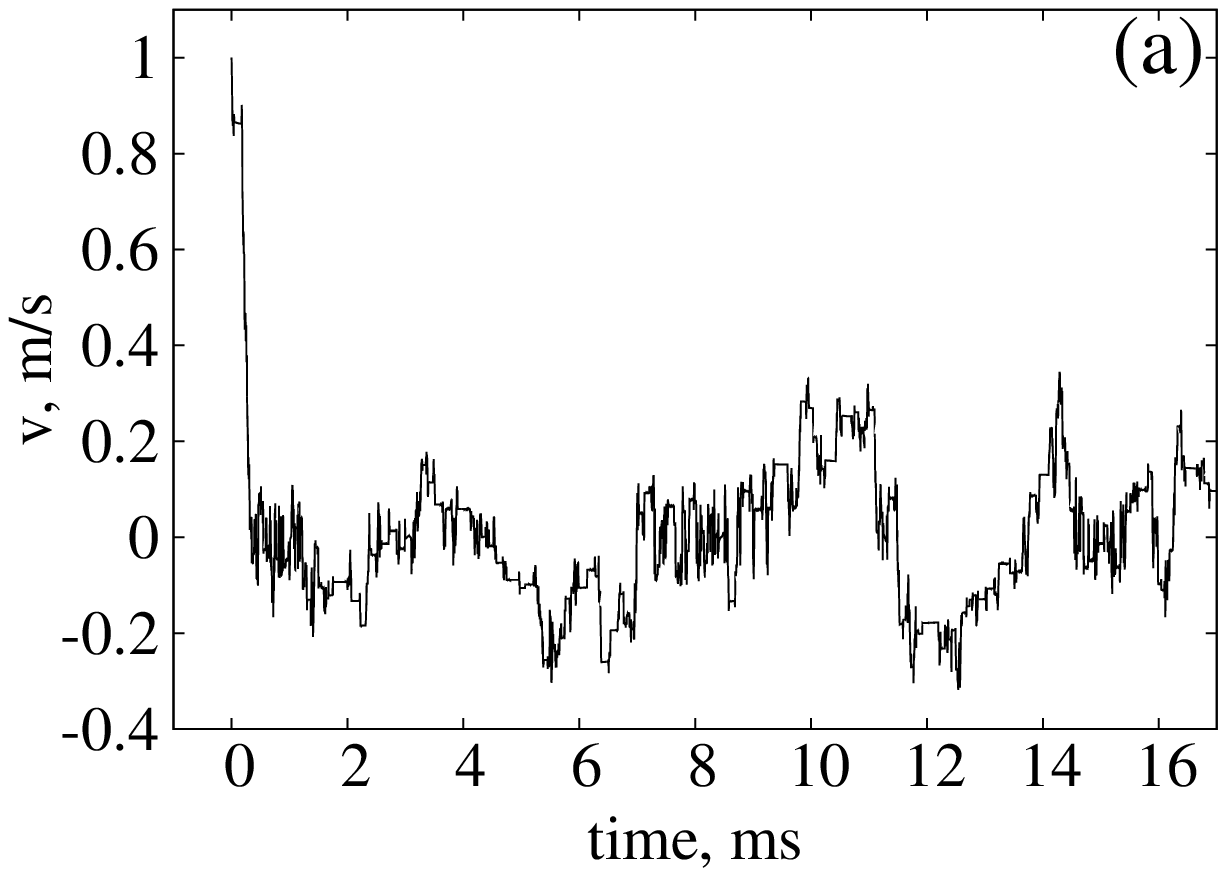}
\bigskip

\includegraphics[width=87mm]{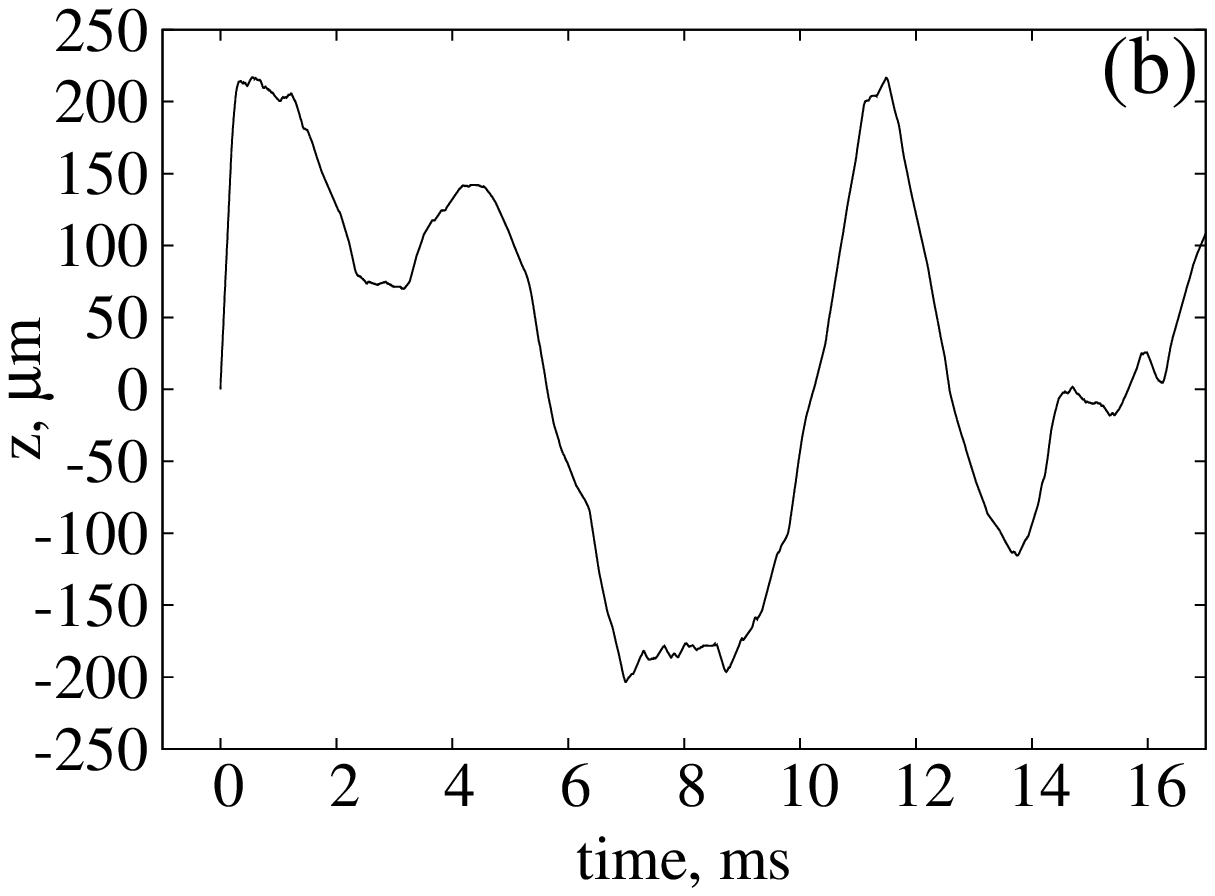}

\caption{An example of SrF molecule motion in the field of the counter-propagating sequences of light pulses: 
(a) $ v(t) $ and  (b) $ z(t) $. Parameters: $ \tau=1 $~ps, $ T=23.8 $~ns, $ \vartheta_{1}=0.1\pi $,
$ \vartheta_{2}=0.014\pi $,
$ \gamma=2\pi\times7 $~MHz, $ \delta_{1}=2\pi\times3.5$~MHz, $ \delta_{2}=2\pi\times7$~MHz, 
$ \lambda=663.3 $~nm, $ M=107 $~Da. The molecule starts
at the center of the trap with initial velocity $ v_{0}=1 $~m/s.}%
\label{figure:7}
\end{center}
\end{figure}

Figure~\ref{figure:7}  shows an example of the atom's motion in the field of the counter-propagating sequences 
of 1-ps light pulses.
Very quickly (0.34~ms after the beginning of the interaction with the field) the atom slows down to zero velocity
and then its velocity fluctuates in the region $ \pm 0.4$~m/s. The atom returns to the center of the trap
approximately after 5.6~ms and then fluctuates in the region $ \pm 0.25$~mm. 

Sometimes, in 2\% cases, the excited molecule relax to $ g_2 $ state.
As a result, we see in Fig.~\ref{figure:7} several almost horizontal pieces.
These pieces corresponds to staying the molecule in the state $ g_2 $, 
where the interaction of the molecule with the field 
is much weaker than in the state $ g_1 $. Between these pieces the velocity time dependence 
resembles one of the two-level atom in the field of the counter-propagating pulse trains, shown
in Fig.~\ref{figure:3}(a). 
The capture range of the trap for the parameters specified in Fig.~\ref{figure:7} 
extends at least from $v=-12 $~m/s to $v= +12 $~m/s.

\begin{figure}[ht]
\begin{center}
\includegraphics[width=87mm]{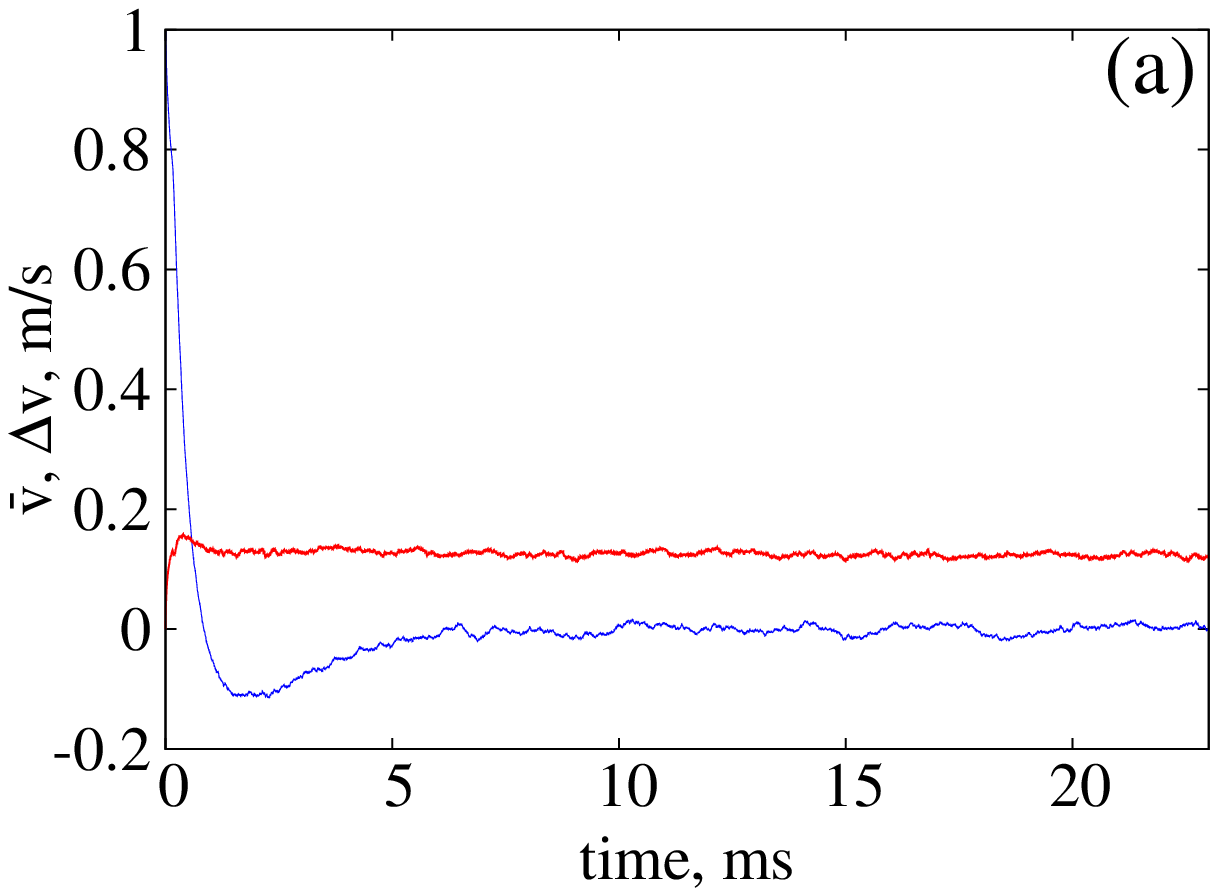}
\bigskip

\includegraphics[width=87mm]{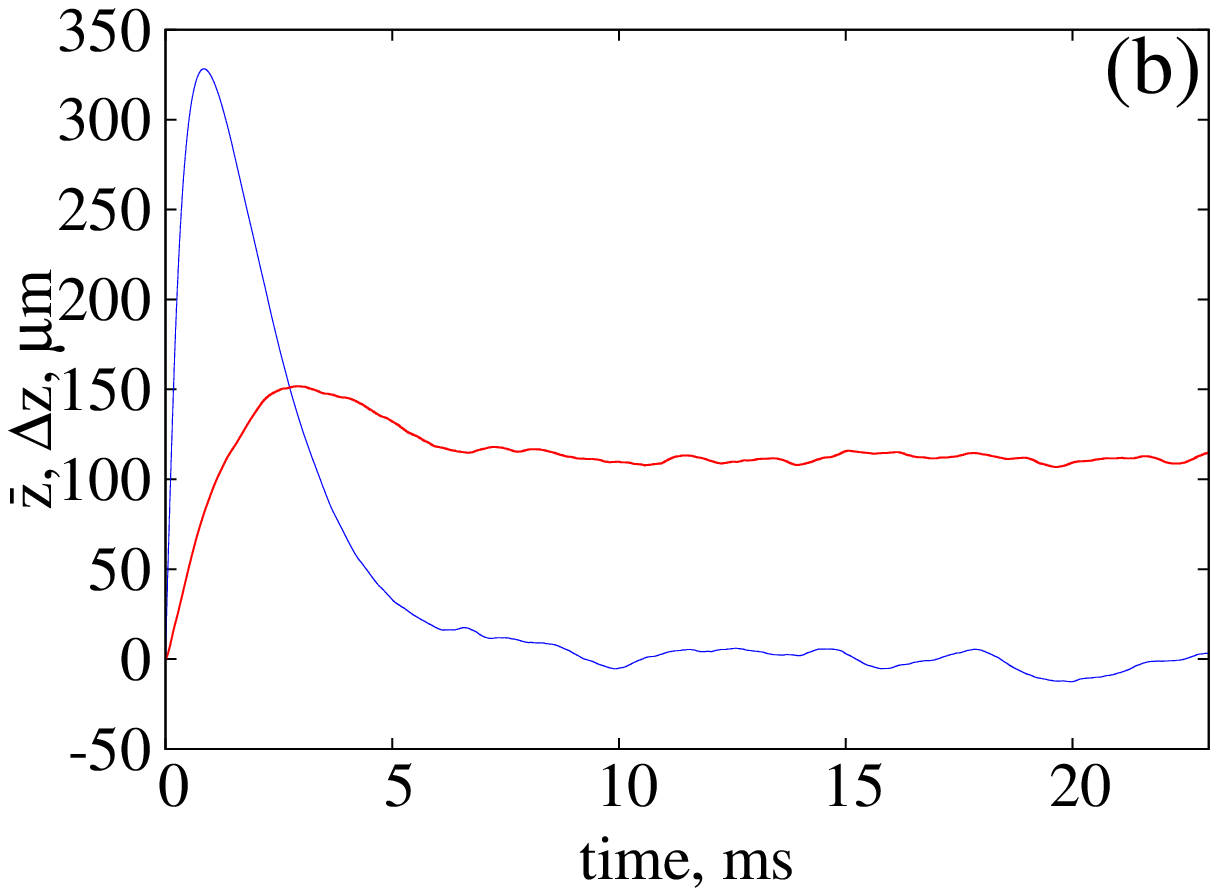}

\caption{The time dependences of average coordinate $ \bar{z}= \left \langle z \right \rangle $ and velocity 
$ \bar{v}= \left \langle v \right \rangle  $,
$ \Delta v=\sqrt{\left \langle v^{2} \right \rangle-\left \langle v \right \rangle^{2}}$ and
$ \Delta z $ for an ensemble of 400 molecules. The parameters are the same of Fig.~\ref{figure:7}.
}%
\label{figure:8}
\end{center}
\end{figure}

The time dependences of average coordinate $ \bar{z}= \left \langle z \right \rangle $, average velocity 
$ \bar{v}= \left \langle v \right \rangle  $ and
$ \Delta v=\sqrt{\left \langle v^{2} \right \rangle-\left \langle v \right \rangle^{2}}$,
$ \Delta z $ for an ensemble of 400 molecules are depicted in 
Fig.~\ref{figure:8}. 
Approximately after 10~$ \mu $s the ensemble of molecules with equal initial 
velocity become localized in the vicinity of the coordinate origin with $ \Delta z=112\,\mu$m
and $ \Delta v =12.2$~cm/s, that slightly larger then $11.4$~cm/s, corresponding to $ T_{\min}=168\, \mu$K.

\subsection{Perspective for the nanoparticle light pulses's trap}

\begin{figure}[ht]
\begin{center}
\includegraphics[width=87mm]{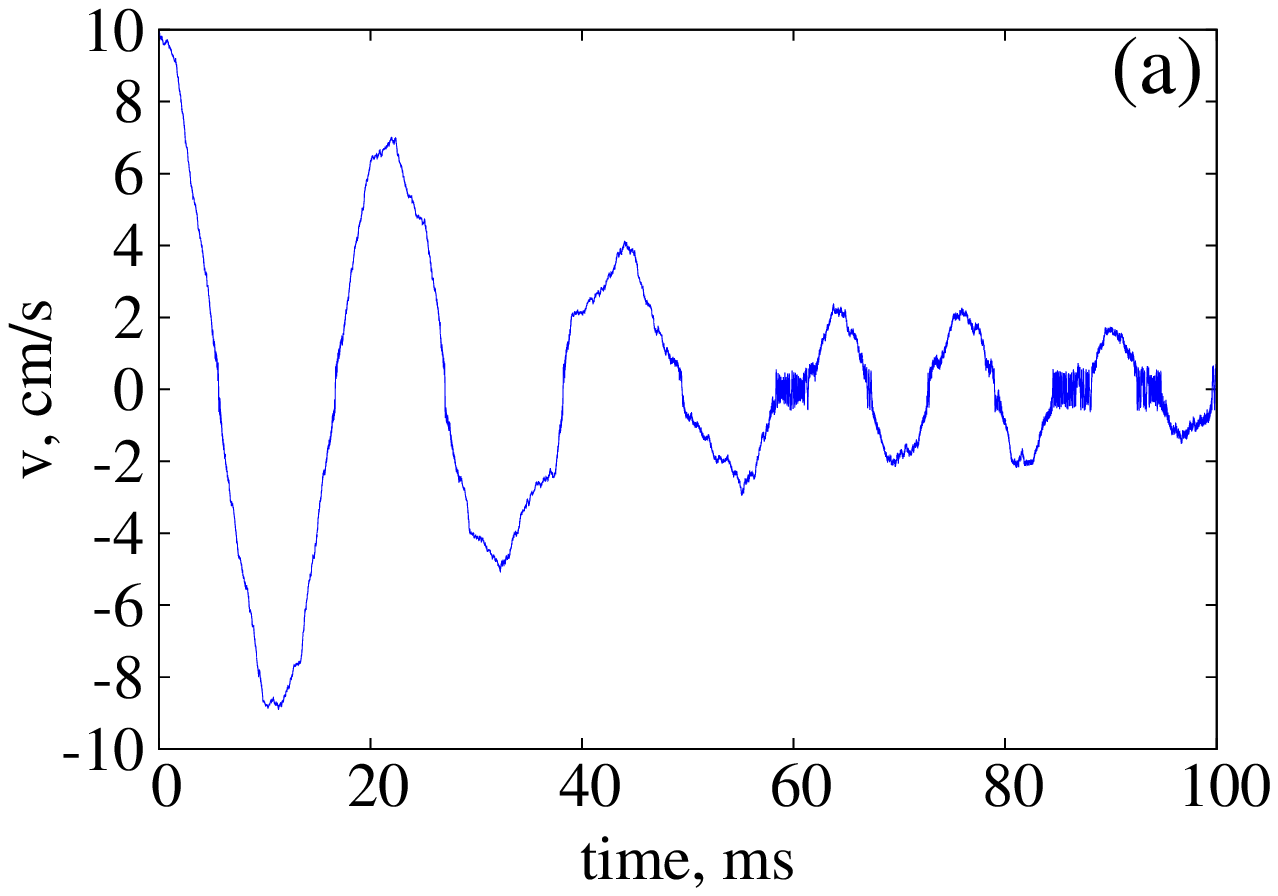}
\bigskip

\includegraphics[width=87mm]{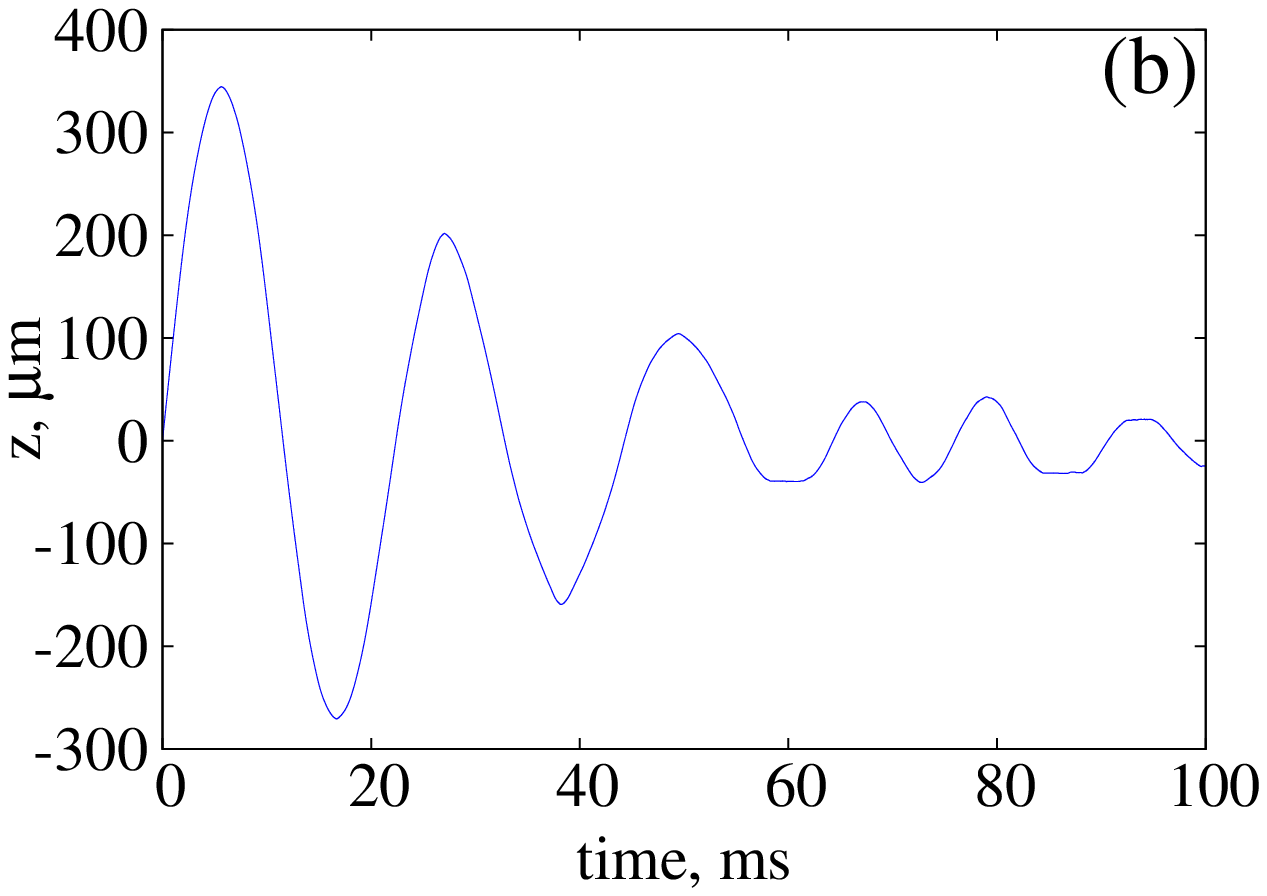}

\caption{An example of a nanoparticle's motion in the field of the counter-propagating sequences of light pulses:
(a) $ v(t) $ and  (b) $ z(t) $. It is supposed that the mass of nanoparticle per every ``active''atom is 
30000~Da.
Parameters: $ \tau=1 $~ps, $ T=10 $~ns, $ \vartheta=0.1\pi $,
$ \gamma=2\pi\times10 $~MHz, $ \delta=2\pi\times5 $~MHz, $ \lambda=600 $~nm. 
The nanoparticle starts at the center of the trap with initial velocity $ v_{0}=10 $~cm/s.}%
\label{figure:9}
\end{center}
\end{figure}

Let's suppose that a nanoparticle includes ``active atoms'' which energetic levels almost are not 
perturbed by the interaction with neighbor atoms (for example, rare earth atoms). We can estimate behavior of the nanoparticle 
in the field of the counter-propagating pulses analyzing the motion of the hypothetical two-level atom
with mass equal to $ M= M_{np}/N_{a}$, where $M_{np}$ is the mass of nanoparticle, $N_{a}$ is 
the number of ``active atoms''. Figure~\ref{figure:9} shows an example of 
a nanoparticle's motion in the field of the counter-propagating sequences of light pulses. 
The pulse's propagation direction is normal to the gravity acceleration. 
As in the case of a sodium atom, the nanoparticle oscillates around the coordinate origin,
where the counter-propagating pulses ``collide''. Amplitude of the oscillations decays in the case 
$ \delta>0 $. Sometimes the nanoparticle oscillates in the vicinity of the field's nodes, that
can be seen in Fig.~\ref{figure:9} (for example, at $ -30.6\, \mu$m, $ -30.9\, \mu$m $ -31.5\, \mu$m,
$-31.8\, \mu$m), jumping from one node to another neighboring node. The period of such oscillation is $ \sim 160\,\mu $s.
The results of calculations shows the favorable perspective for experimental realization of the trap
for nanoparticles with included ``active'' atoms.

\section{Conclusions}
\label{sec-conc}
We simulated atomic and molecular motion (one particle and ensemble of particles)
in the field of weak counter-propagating light pulses and showed, that these pulses form
a light trap which, beside trapping of particles, cool them down to the Doppler temperature limit.
Analyzing atoms, we used the two-level model of the atom-field interaction. The molecules in the trap
were analyzed in the approximation of the three-level $ \Lambda $-type model, which can be applicable
for the molecules with almost diagonal Frank-Condon factor arrays. 
The parameter of the atom-field interaction in the case of molecules must eliminate the two-photon resonance condition.
Velocity capture range for atoms and molecules exceeds 10~m/s,
spatial capture range is about 100~$ \mu\mathrm{m} $.

We also discussed the applicability of the trap to confinement of nanoparticle, assuming the nanoparticles
includes ``active'' atoms, i.e. atoms with transitions close to carrier frequency of the pulses. 
The simulation result shows the good perspective of the realization of such a trap.

\section*{ACKNOWLEDGMENTS}

This research was supported by the State goal-oriented scientific and engineering program "Nanotechnologies and Nanomaterials" (1.1.4.13/14-H25) and by the State Fund for Fundamental Researches of Ukraine (project F53.2/001).

\bibliography{Pulse-trap-cooling-arxiv}
\end{document}